\documentstyle [12pt,epsf]{article}

\input{epsf}
\begin{document}
\begin{titlepage}
\begin{center}

 \vspace{-0.7in}

{\large \bf Stochastic Quantization of $(\lambda\varphi^{4})_d$ Scalar Theory:\\
Generalized Langevin Equation with Memory Kernel}\\
 \vspace{.3in}{\large\em G. Menezes\,\footnotemark[1]
and N. F. Svaiter
\footnotemark[2]}\\
Centro Brasileiro de Pesquisas F\'{\i}sicas\,-CBPF,\\
 Rua Dr. Xavier Sigaud 150,\\
 Rio de Janeiro, RJ, 22290-180, Brazil \\

\subsection*{\\Abstract}
\end{center}
\baselineskip .18in

We review the method of stochastic quantization for a scalar field
theory. We first give a brief survey for the case of
self-interacting scalar fields, implementing the stochastic
perturbation theory up to the one-loop level. The divergences
therein are taken care of by employing the usual prescription of
the stochastic regularization, introducing a colored random noise
in the Einstein relations. We then extend this formalism to the
case where we assume a Langevin equation with a memory kernel. We
have shown that, if we also maintain the Einstein's relations with
a colored
noise, there is convergence to a non-regularized theory.\\

PACS numbers: 03.70+k, 04.62.+v

\footnotetext[1]{e-mail:\,\,gsm@cbpf.br}
\footnotetext[2]{e-mail:\,\,nfuxsvai@cbpf.br}

\end{titlepage}
\newpage\baselineskip .20
in
\section{Introduction}
$\,\,\,\,\,\,\,$

In the last century Parisi and Wu introduced the stochastic
quantization \cite{parisi}. This new quantization method differs
from the others, the canonical and the path integral field
quantization, based in the Hamiltonian and the Lagrangian,
respectively, in many aspects. The method starts from a classical
equation of motion, but not from Hamiltonian or Lagrangian, and
consequently can be used to quantize dynamical systems without
canonical formalism. Furthermore, it is useful in situations where
the others methods lead to difficult problems and can bring us new
important results. As stressed by Rumpf \cite{rumpf}, since the
stochastic quantization is quite different from the other
quantization methods, it can reveal new structural elements of the
classical theory which so far have gone unnoticed. The main idea
of the stochastic quantization is that a $d$-dimensional quantum
system is equivalent to a $(d+1)$-dimensional classical system
with random fluctuations. Some of the most important papers in the
subject can be found in Ref. \cite{ii}. A brief introduction to
the stochastic quantization can be found in the Ref.
\cite{namiki1} and Ref. \cite{sakita}. See also the Ref.
\cite{damre}.

In this paper we would like to discuss the stochastic quantization
method for the $(\lambda\varphi^{4})_d$ scalar theory using a
Langevin equation with a memory kernel and a colored noise. We
inquiry which kind of theory appears in the asymptotic limit of
this non-Markovian process at one-loop level. Does this
stationary, Gaussian, non-Markovian theory reaches the equilibrium
with the structure of its ultraviolet divergences under control?

The method of stochastic quantization can be summarized by the
following steps. First, starting from a field defined in Minkowski
spacetime, after analytic continuation to imaginary time, the
Euclidean counterpart, i.e., the field living in an Euclidean
space, is obtained. Second, it is introduced a monotonically
crescent Markov parameter, called in the literature "fictitious
time" and also a random noise field $\eta(\tau,x)$, which
simulates the coupling between the classical system and a heat
reservoir. It is assumed that the fields defined at the beginning
in a $d$-dimensional Euclidean space also depends on the Markov
parameter, therefore the field and a random noise field are
defined in a $(d+1)$-dimensional manifold. One starts with the
system out of equilibrium at an arbitrary initial state. It is
then forced into equilibrium assuming that its evolution is
governed by a Markovian Langevin equation with a white random
noise field. In fact, this evolution is described by a process
which is stationary, Gaussian and Markovian. Finally, the
$n$-point correlation functions of the theory in the
$(d+1)$-dimensional space are defined by performing averages over
the random noise field with a Gaussian distribution, that is,
performing the stochastic averages
$\langle\,\varphi(\tau_{1},x_{1})
\varphi(\tau_{2},x_{2})...\varphi(\tau_{n},x_{n})\,\rangle_{\eta}$.
The $n$-point Schwinger functions of the Euclidean $d$-dimensional
theory are obtained  evaluating these $n$-point stochastic
averages $\langle\,\varphi(\tau_{1},x_{1})
\varphi(\tau_{2},x_{2})...\varphi(\tau_{n},x_{n})\,\rangle_{\eta}$
when the Markov parameter goes to infinity
$(\tau\rightarrow\infty)$, and the equilibrium is reached. This
can be proved in different ways for the particular case of
Euclidean scalar field theory. One can use, for instance, the
Fokker-Planck equation \cite{fp} \cite{ili} associated with the
equations describing the stochastic dynamic of the system. A
diagrammatical technique \cite{grimus} has also been used to prove
such equivalence.

The original method proposed by Parisi and Wu was extended to
include process with fermions \cite{f1} \cite{f2} \cite{f3}. The
first question that appears in this context is if make sense the
Brownian problem with anticommutating numbers. It can be shown
that, for massless fields, there will not be a convergence factor
after integrating the Markovian Langevin equation. Therefore the
equilibrium is not reached. One way of avoiding this problem is to
introduce a kernel in the Langevin equation describing the
evolution of two Grassmannian fields.

Usually, the Parisi-Wu schemes of quantization applied to bosonic
and fermionic fields converge towards non-regularized theories. In
order to obtain finite results, the original stochastic process is
modified to a non-Markovian process, taking into account a colored
noise in the Einstein relations \cite{iengo} \cite{halp}. In few
words, the idea of the stochastic regularization is to start from
an interacting theory, then construct Langevin tree-graphs, where
each leg of which ends in a regularized noise factor. Since it is
possible to obtain the loops of the theory by contracting the
noise factors, one ends up with a theory where every closed loop
contains at least some power of the regulator. With this
modification, one can show that the system converges towards a
regularized theory. The next step to construct a finite theory is
to use for example a minimal-subtraction scheme in which the
ultraviolet divergent contributions are eliminated.

However, different types of non-Markovian processes are allowed in
principle; for instance, we can modify the Langevin equation
introducing a memory kernel. To make sure that the generalized
Langevin equation can also be used as a quantization tool, one
must check that the process converges in the asymptotic limit and
also that converges to the correct equilibrium distribution. In
this paper we address the following question: which kind of
interacting field theory appears in the asymptotic limit of this
non-Markovian process at one-loop level? We have shown that
although a system with a stationary, Gaussian, non-Markovian
Langevin equation with a memory kernel and a colored noise
converges to equilibrium, we obtain a non-regularized theory.

The organization of the paper is the following: in section 2 we
briefly discuss the Markovian and the non-Markovian Langevin
equation in stochastic processes. We discuss in section 3 the
stochastic quantization for the free scalar theory. The stochastic
quantization for the $(\lambda\varphi^{4})_{d}$ self-interaction
scalar theory with the usual stochastic regularization method is
presented in section 4. In section 5 assuming a Langevin equation
with a memory kernel, we study the free and the interacting field
theory that appears in the asymptotic limit of the non-Markovian
process. In section 6, we developed a Fokker-Planck approach in
order to understand the results obtained in the previous section.
Conclusions are given in section 7. To simplify the calculations
we assume the units to be such that $\hbar=c=k_{B}=1$.
\newpage

\section{Markovian and non-Markovian Langevin equation}\

Let us briefly discuss the role of the Langevin equation in
non-equilibrium statistical mechanics \cite{z}. One way to treat
the dynamics of non-equilibrium systems is to use the theory of
Brownian motion. The fundamental equation is the Langevin
equation, and it contains both frictional and random forces. The
equation of motion for a Brownian particle is given by
\begin{equation}
m\frac{d}{dt}{\bf v}(t)=-\xi\,{\bf v}(t)+\delta {\bf F}(t),
\label{1}
\end{equation}
where $-\xi\,{\bf v}(t)$ is the frictional force and $\delta\,{\bf
F}(t)$ is the stochastic or fluctuating force, also called the
random force. In Eq.(\ref{1}), $m$ is the mass of the Brownian
particle immersed in the fluid. Note that the total force that
acts on each  Brownian particle has been partioned into a friction
part and also a fluctuating part. The effects of the fluctuation
force can be summarized by given the Einstein relations, namely,
\begin{equation}
\langle\,\delta {\bf F}(t)\,\rangle=0
\label{3}
\end{equation}
and also
\begin{equation}
\langle\,\delta\,F^{i}(t)\,\delta F^{j}(t')\,
\rangle=2B\delta_{ij}\delta(t-t'),
\label{4}
\end{equation}
where $B$ is the measure of the strength of the fluctuating force.
The Eq.(\ref{3}) tell us that the random force is zero on average
and the delta function in Eq.(\ref{4}) indicates that there is no
correlation between impacts in any two distinct time intervals.
Since the frictional force depends just on the velocity of the
particle and not on its earlier values, we are describing a
Markovian process. The Langevin equation can be solved to give
\begin{equation}
{\bf v}(t)=\exp\left(-\frac{\xi }{m}\,t\right){\bf
v}(0)+\frac{1}{m}
\int_{0}^{\,t}ds\exp\left(-\frac{\xi}{m}(t-s)\right)\delta {\bf
F}(s). \label{2}
\end{equation}
Using the fact that the fluctuating force has a Gaussian distribution
defined by its first and seconds moments,
it is possible to show that the mean squared velocity
of each Brownian particle is
\begin{equation}
\langle\,v^{2}(t)\,\rangle=\exp\left(-\frac{2\xi
}{m}\,t\right)v^{2}(0)+
\frac{B}{\xi\,m}\left(1-\exp\left(-\frac{2\xi}{m}\right)t\right),
\label{5}
\end{equation}
where the brackets $\langle...\rangle$ in the
Eq. (\ref{5}) denote stochastic averaging.

There are two points that we would like to stress. First, it is
assumed that the fluctuating force $\delta {\bf F}(t)$ has a
Gaussian distribution. Therefore it is possible to relate the
strength $B$ of the random noise to the magnitude $\xi$ of the
friction $(B=\xi T)$, where $T$ is the temperature of the bath.
This is the well known fluctuation-dissipation theorem. Note that
the Langevin equation considered in Eq.(\ref{1}) is called
Markovian, since the frictional force $(-\xi {\bf v}(t))$ at $t$
is proportional to the velocity at $t$ and the noise is
delta-function correlated or white. Being more specific, the
Fourier transform of the correlation function of the noise, that
is, the spectral density, is independent of frequency.

The Langevin equation may be generalized to non-Markovian types
assuming that the friction at time $t$ can depend on the history
of the velocity ${\bf v}(s)$ for times $s$ that are earlier than
$t$ \cite{fox1} \cite{fox2} \cite{kubo}. In this situation, the
friction coefficient must be replaced by a memory function. The
Langevin equation with a dissipative memory kernel, that we will
call a generalized Langevin equation, reads
\begin{equation}
m\frac{d}{dt}{\bf v}(t)=-\int_{0}^{\infty}ds\,M(s){\bf v}(t-s)+\delta
{\bf F}(t).
\label{7}
\end{equation}
In the paper we will assume a non-Markovian Langevin equation with
a memory kernel and a colored noise. We will examine the
interacting scalar field theory that appears in the asymptotic
limit of this non-Markovian process. To bring the reader to this
new approach let us first consider the standard stochastic
quantization of a free scalar field theory.

\section{Stochastic quantization for a
free scalar theory}\

Let us consider a neutral scalar field with a
$(\lambda\varphi^{4})$ self-interaction, defined in a
$d$-dimensional Minkowski spacetime. The vacuum persistence
functional is the generating functional of all vacuum expectation
value of time-ordered products of the theory. The Euclidean field
theory can be obtained by analytic continuation to imaginary time
supported by the positive energy condition for the relativistic
field theory. In the Euclidean field theory, we have the Euclidean
counterpart for the vacuum persistence functional, that is, the
generating functional of complete Schwinger functions. Actually,
the $(\lambda\varphi^{4})_{d}$ Euclidean theory is defined by
these Euclidean Green's functions. The Euclidean generating
functional $Z[h]$ is formally defined by the following functional
integral:
\begin{equation}
Z[h]=\int [d\varphi]\,\, \exp\left(-S_{0}-S_{I}+ \int d^{d}x\,
h(x)\varphi(x)\right), \label{8}
\end{equation}
where the action that usually describes a free scalar field is
\begin{equation}
S_{0}[\varphi]=\int d^{d}x\, \left(\frac{1}{2}
(\partial\varphi)^{2}+\frac{1}{2}
m_{0}^{2}\,\varphi^{2}(x)\right), \label{9}
\end{equation}
and the interacting part, defined by the non-Gaussian contribution,
is
\begin{equation}
S_{I}[\varphi]= \int d^{d}x\,\frac{\lambda}{4!} \,\varphi^{4}(x).
\label{10}
\end{equation}
In Eq.(\ref{8}), $[d\varphi]$ is a translational invariant
measure, formally given by $[d\varphi]=\prod_{x} d\varphi(x)$. The
terms $\lambda$ and $m_{0}^{2}$ are respectively the bare coupling
constant and the squared mass of the model. Finally, $h(x)$ is a
smooth function that we introduce to generate the Schwinger
functions of the theory by functional derivatives. In the
weak-coupling perturbative expansion, which is the conventional
procedure, we perform a formal perturbative expansion with respect
to the non-Gaussian terms of the action. As a consequence of this
formal expansion, all the $n$-point unrenormalized Schwinger
functions are expressed in a powers series of the bare coupling
constant $\lambda$ \cite{livron} \cite{id}.

The aim of this section is to discuss the stochastic quantization
of a free scalar field. It can be shown that it is equivalent to
the usual path integral quantization. The starting point of the
stochastic quantization to obtain the Euclidean field theory is a
Markovian Langevin equation. Assume an Euclidean $d$-dimensional
manifold, where we are choosing periodic boundary conditions for a
scalar field and also a random noise. In other words, they are
defined in a $d$-torus $\Omega\equiv\,T\,^d$. To implement the
stochastic quantization we supplement the scalar field
$\varphi(x)$ and the random noise $\eta(x)$ with an extra
coordinate $\tau$, the Markov parameter, such that
$\varphi(x)\rightarrow \varphi(\tau,x)$ and $\eta(x)\rightarrow
\eta(\tau,x)$. Therefore, the fields and the random noise are
defined in a domain: $T\,^{d}\times R\,^{(+)}$. Let us consider
that this dynamical system is out of equilibrium, being described
by the following equation of evolution:
\begin{equation}
\frac{\partial}{\partial\tau}\varphi(\tau,x)
=-\frac{\delta\,S_0}{\delta\,\varphi(x)}|_{\varphi(x)=\varphi(\tau,\,x)}+
\eta(\tau,x),
\label{23}
\end{equation}
where $\tau$ is a Markov parameter, $\eta(\tau,x)$ is a random
noise field and $S_0$ is the usual free action defined in
Eq.(\ref{9}). For a free scalar field, the Langevin equation reads
\begin{equation}
\frac{\partial}{\partial\tau}\varphi(\tau,x)
=-(-\Delta+m^{2}_{0}\,)\varphi(\tau,x)+ \eta(\tau,x),
 \label{24}
\end{equation}
where $\Delta$ is the $d$-dimensional Laplace operator. The
Eq.(\ref{24}) describes a Ornstein-Uhlenbeck process and we are
assuming the Einstein relations, that is:
\begin{equation}
\langle\,\eta(\tau,x)\,\rangle_{\eta}=0,
\label{28}
\end{equation}
and for the two-point correlation function associated with the
random noise field
\begin{equation}
\langle\, \eta(\tau,x)\,\eta(\tau',x')\,
\rangle_{\eta}\,=2\delta(\tau-\tau')\,\delta^{d}(x-x'), \label{29}
\end{equation}
where $\langle\,...\rangle_{\eta}$ means stochastic averages. In a
generic way, the stochastic average for any functional of
$\varphi$ given by $F[\varphi\,]$ is defined by
\begin{equation}
\langle\,F[\varphi\,]\,\rangle_{\eta}=
\frac{\int\,[d\eta]F[\varphi\,]\exp\biggl[-\frac{1}{4} \int d^{d}x
\int d\tau\,\eta^{2}(\tau,x)\bigg]}
{\int\,[d\eta]\exp\biggl[-\frac{1}{4} \int d^{d}x \int
d\tau\,\eta^{2}(\tau,x)\bigg]}. \label{36}
\end{equation}
Let us define the retarded Green function for the diffusion
problem that we call $G(\tau-\tau',x-x')$. The retarded Green
function satisfies $G(\tau-\tau',x-x')=0$ if $\tau-\tau'<0$ and
also
\begin{equation}
\Biggl[\frac{\partial}{\partial\tau}+(-\Delta_{x}+m^{2}_{0}\,)
\Bigg]G(\tau-\tau',x-x')=\delta^{d}(x-x')\delta(\tau-\tau').
\label{25}
\end{equation}
Using the retarded Green function and the initial condition
$\varphi(\tau,x)|_{\tau=0}=0$, the solution for Eq.(\ref{24})
reads
\begin{equation}
\varphi(\tau,x)=\int_{0}^{\tau}d\tau'\int_{\Omega}d^{d}x'\,
G(\tau-\tau',x-x')\eta(\tau',x').
\label{26}
\end{equation}
In the following we are interested in calculating the quantity
$\langle\,\varphi(\tau,x)\varphi(\tau',x')\, \rangle_{\eta}$.
Using Eq.(\ref{28}), Eq.(\ref{29}) and  Eq.(\ref{26}), we have
\begin{equation}
\langle\,\varphi
(\tau_{1},x_{1})\varphi(\tau_{2},x_{2})\,\rangle_{\eta}=
2\int_{0}^{\,min(\tau_{1},\tau_{2})}
d\tau'\int_{\Omega}d^{d}x'G(\tau_{1}-\tau',x_{1}-x')\,G(\tau_{2}-\tau',
x_{2}-x'),
 \label{31}
\end{equation}
where $min(\tau_{1},\tau_{2})$ means the minimum of $\tau_{1}$ and
$\tau_{2}$.
Using a Fourier representation, the two-point correlation function
$\langle\,\varphi(\tau,x)\varphi(\tau',x')\,\rangle_{\eta}\equiv
D(\tau,x;\tau',x')$ is given by
\begin{equation}
D(\tau,x;\tau',x')= \frac{1}{(2\pi)^{d}}\int\,d^{d}p \,\frac{\,\,
e^{-ip(x-x')}}{(p^{2}+m_{0}^{2})}\,e^{-(p^{2}+m_{0}^{2})(\tau-\tau')}.
\label{n1}
\end{equation}
It is not difficult to show that Eq.(\ref{n1}) can be written as:
\begin{equation}
D(\tau,x;\tau',x') = \frac{1}{(2\pi)\,^{d}}
\sum_{n=0}^{\infty}\frac{(-1)^{n}}{\Gamma(1-n)n!}(\tau-\tau')^{n}
\left(\frac{m_{0}}{r}\right)^{\frac{d}{2}+n-1}
K_{\frac{d}{2}+n-1}(m_{0}\,r). \label{asa}
\end{equation}
where $r = \mid x-x'\mid$ and $K_{\nu}$ is the modified Bessel
function of order $\nu$.

 We can use the Fourier analysis to show that when the Markov
parameters $\tau$ and $\tau'$ go to infinity we recover the
standard Euclidean free field theory. Therefore let us define the
Fourier transforms for the field  and the noise given by
$\varphi(\tau,k)$ and $\eta(\tau,k)$. We have respectively
\begin{equation}
\varphi(\tau,k)=\frac{1}{(2\pi)^\frac{d}{2}} \int\,d^{d}x\,
e^{-ikx}\,\varphi(\tau,x), \label{33}
\end{equation}
and
\begin{equation}
\eta(\tau,k)=\frac{1}{(2\pi)^\frac{d}{2}} \int\,d^{d}x\,
e^{-ikx}\,\eta(\tau,x). \label{34}
\end{equation}
Substituting Eq.(\ref{33}) in Eq.(\ref{9}), the free action for
the scalar field in the $(d+1)$-dimensional space writing in terms
of the Fourier coefficients reads
\begin{equation}
S_{0}[\varphi(k)]\,|_{\varphi(k)=\varphi(\tau,\,k)}=
\frac{1}{2}\int\,d^{d}k\,\varphi(\tau,k)(k^{2}+m_{0}^{2})\varphi(\tau,k).
\label{35}
\end{equation}
Substituting Eq.(\ref{33}) and Eq.(\ref{34}) in Eq.(\ref{24}) we
have that each Fourier coefficient satisfies a Langevin equation
given by
\begin{equation}
\frac{\partial}{\partial\tau}\varphi(\tau,k)
=-(k^{2}+m^{2}_{0})\varphi(\tau,k)+
\eta(\tau,k).
\label{36}
\end{equation}
The solution for this equation reads
\begin{equation}
\varphi(\tau,k)=\exp\left(-(k^{2}+m_{0}^{2})\tau\right)\varphi(0,k)+
\int_{0}^{\tau}d\tau'\exp\left(-(k^{2}+m_{0}^{2})(\tau-\tau')\right)
\eta(\tau',k).
\label{37}
\end{equation}
Using the Einstein relation, we get that the Fourier coefficients
for the random noise satisfies
\begin{equation}
\langle\,\eta(\tau,k)\,\rangle_{\eta}=0 \label{38}
\end{equation}
and
\begin{equation}
\langle\,\eta(\tau,k)\eta(\tau',k')\,\rangle
_{\eta}=2(2\pi)^{d}\delta(\tau-\tau')\,\delta^{d}(k+k').
\label{39}
\end{equation}
Before investigate the interacting field theory, let us calculate
the Fourier representation for the two-point correlation function,
i.e., $\langle\,\varphi(\tau,k)\varphi(\tau',k')\,\rangle_{\eta}$.
Using Eq.(\ref{37}), we obtain three contributions to the scalar
two-point correlation function. The first one is given by
\begin{equation}
\exp\left(-(k^{2}+m_{0}^{2})\,\tau+(k'^{2}+m_{0}^{2})\,\tau'\right)
\varphi(0,k)\varphi(0,k'), \label{40}
\end{equation}
and decay to zero at long time. Let us assume that
$\varphi(\tau,k)|_{\tau=0}=0$. There are also two crossed terms,
each first order in the noise Fourier component given by
\begin{equation}
2\,\varphi(0,k)\exp\left(-(k^{2}+m_{0}^{2})\tau\right)
\int_{0}^{\tau'}ds\, \exp\left(-(k'^{2}+m_{0}^{2})(\tau'-s)\right)
\eta(s,k'). \label{41}
\end{equation}
Since we are assuming the Einstein relations, i.e.,
$\langle\,\eta(\tau,x)\,\rangle _{\eta}=0$, on averaging on noise,
these cross terms vanish. The final term is second-order in the
noise Fourier component. Again, the solution subject to the
initial condition $\varphi(\tau,k)|_{\tau=0}=0$ can be used to
give
\begin{equation}
\int_{0}^{\tau}ds\, \exp\left(-(k^{2}+m_{0}^{2})(\tau-s)\right)
\eta(s,k)\int_{0}^{\tau'}d\sigma
\exp\left(-(k'^{2}+m_{0}^{2})(\tau'-\sigma)\right)
\eta(\sigma,k'). \label{42}
\end{equation}
Again averaging on noise and using the Einstein relation given by
Eq.(\ref{39}) we have that this term becomes
\begin{equation}
2\delta^{d}(k+k')\int^{\,min(\tau,\tau')}_{0}\, ds
\exp\left(-(k^{2}+m_{0}) (\tau+\tau'-2s)\right). \label{43}
\end{equation}
Assuming that $\tau=\tau'$ and  using
$\langle\,\varphi(\tau,k)\varphi(\tau',k')\,\rangle_{\eta}|_{\tau=\tau'}\equiv
D(k,k';\tau,\tau')$ we have
\begin{equation}
D(k;\tau,\tau)=(2\pi)^d\delta^{d}(k+k')\frac{1}{(k^{2}+m_{0}^{2})}\left(1-\exp\left(-2\tau
(k^{2}+m_{0}^{2})\right)\right).
\label{44}
\end{equation}
In the following, we are redefining the two-point correlation
function as $D(k;\tau,\tau)\rightarrow(2\pi)^{d}D(k;\tau,\tau)$.
In the limit when $\tau\rightarrow\infty$ we recover the standard
two-point function of the Euclidean free field theory. Before
going to the next section, we would like to mention the existence
of more general Markovian Langevin equations. We can introduce a
kernel defined in the $d$-torus. The kerneled Langevin equation
reads:
\begin{equation}
\frac{\partial}{\partial\tau}\varphi(\tau,x) =-\int
d^dy\,K(x,y)\frac{\delta\,S_0}{\delta\,\varphi(y)}|_{\varphi(y)=\varphi(\tau,\,y)}+
\eta(\tau,x). \label{23}
\end{equation}
The second moment of the noise field will be modified to:
\begin{equation}
\langle\, \eta(\tau,x)\,\eta(\tau',x')\,
\rangle_{\eta}\,=2\delta(\tau-\tau')\,K(x,x'). \label{29b}
\end{equation}
Choosing an appropriate kernel, it can be shown that all the above
conclusions remain unchanged.

\section{Stochastic quantization for the
$(\lambda\varphi^{4})_{d}$ scalar theory}

In this section we will analyze the stochastic quantization for
the $(\lambda\varphi^{4})_{d}$ self-interaction scalar theory. In
this case the Langevin equation reads
\begin{equation}
\frac{\partial}{\partial\tau}\varphi(\tau,x)
=-(-\Delta+m^{2}_{0}\,)\varphi(\tau,x)-\frac{\lambda}{3!}\varphi^{3}(\tau,x)+
\eta(\tau,x). \label{35}
\end{equation}
The two-point correlation function associated with the random
field is given by the Einstein relation, while the other connected
correlation functions vanish, i.e.,
\begin{equation}
\langle\,\eta(\tau_{1},x_{1})\eta(\tau_{2},x_{2})...\eta(\tau_{2k-1},
x_{2k-1})\,\rangle_{\eta}=0,
\label{377}
\end{equation}
and also
\begin{equation}
\langle\eta(\tau_{1},x_{1})...\eta(\tau_{2k},x_{2k})\,\rangle_{\eta}=
\sum\,\langle\eta(\tau_{1},x_{1})\eta(\tau_{2},x_{2})\,\rangle_{\eta}
\langle\,\eta(\tau_{k},x_{k})\eta(\tau_{l},x_{l})\,\rangle_{\eta}...,
\label{388}
\end{equation}
where the sum is to be taken over all the different ways in which
the $2k$ labels can be divided into $k$ parts, i.e., into $k$
pairs. Performing Gaussian averages over the white random noise,
it is possible to prove that
\begin{equation}
\lim_{\tau\rightarrow\infty}
\langle\,\varphi(\tau_{1},x_{1})\varphi(\tau_{2},x_{2})...
\varphi(\tau_{n},x_{n}) \,\rangle_{\eta}= \frac{\int
[d\varphi]\varphi(x_{1})\varphi(x_{2})...\varphi(x_{n})
\,e^{-S(\varphi)}} {\int [d\varphi]\,e^{-S(\varphi)}}, \label{399}
\end{equation}
where $S(\varphi)=S_0(\varphi)+S_{I}(\varphi) $ is the
$d$-dimensional action. This result leads us to consider the
Euclidean path integral measure a stationary distribution of a
stochastic process. Note that the solution of the Langevin
equation needs a given initial condition. As for example
\begin{equation}
\varphi(\tau,x)|_{\tau=0}=\varphi_{0}(x).
\label{40}
\end{equation}

Let us use the Langevin equation to perturbatively solve the
interacting field theory. One way to handle the Eq.(\ref{35}) is
with the method of Green's functions. We defined the retarded
Green function for the diffusion problem in the Eq.(\ref{25}). Let
us assume that the coupling constant is a small quantity.
Therefore to solve the Langevin equation in the case of a
interacting theory we use a perturbative series in $\lambda$.
Therefore we can write
\begin{equation}
\varphi(\tau,x)=\varphi^{(0)}(\tau,x)+\lambda\varphi^{(1)}(\tau,x)+
\lambda^{2}\varphi^{(2)}(\tau,x)+...
\label{41}
\end{equation}
Substituting the Eq.(\ref{41}) in the Eq.(\ref{35}), and if we
equate terms of equal power in $\lambda$, the resulting equations
are
\begin{equation}
\Biggl[\frac{\partial}{\partial\tau}+(-\Delta_{x}+m^{2}_{0}\,)
\Bigg]\varphi^{(0)}(\tau,x)=\eta(\tau,x),
\label{42}
\end{equation}
\begin{equation}
\Biggl[\frac{\partial}{\partial\tau}+(-\Delta_{x}+m^{2}_{0}\,)
\Bigg]\varphi^{(1)}(\tau,x)=-\frac{1}{3!}
\left(\varphi^{(0)}(\tau,x)\right)^{3},
\label{43}
\end{equation}
\begin{equation}
\Biggl[\frac{\partial}{\partial\tau}+(-\Delta_{x}+m^{2}_{0}\,)
\Bigg]\varphi^{(2)}(\tau,x)=-\frac{1}{2!}
\left(\varphi^{(0)}(\tau,x)\right)^{2}\varphi^{(1)}(\tau,x),
\label{43´}
\end{equation}
and so on. Using the retarded Green function and assuming that
$\varphi^{\,(q)}(\tau,x)|_{\tau=0}=0,\,\,\forall\,q$, the solution
to the first equation given by Eq.(\ref{42}) can be written
formally as
\begin{equation}
\varphi^{(0)}(\tau,x)=\int_{0}^{\tau}d\tau'\int_{\Omega}d^{d}x'\,
G(\tau-\tau',x-x')\eta(\tau',x').
\label{sol}
\end{equation}
The second equation given by Eq.(\ref{43}) can also be solved
using the above result. We obtain
\begin{eqnarray}
\varphi^{(1)}(\tau,x)&=&
-\frac{1}{3!}\int_{0}^{\tau}d\tau_{1}\int_{\Omega}d^{d}x_{1}\,
G(\tau-\tau_{1},x-x_{1})\nonumber \\
&&\left(\int_{0}^{\tau_{1}}d\tau'\int_{\Omega}d^{d}x'\,
G(\tau_{1}-\tau',x_{1}-x')\eta(\tau',x') \right)^{3}. \label{44}
\end{eqnarray}
We have seen that we can generate all the tree diagrams with the
noise field contributions. We can also consider the $n$-point
correlation function
$\langle\,\varphi(\tau_{1},x_{1})\varphi(\tau_{2},x_{2})...
\varphi(\tau_{n},x_{n})\,\rangle_{\eta}$. Substituting the above
results in the $n$-point correlation function, and taking the
random averages over the white noise field using the
Wick-decomposition property defined by Eq.(\ref{388}) we generate
the stochastic diagrams. Each of these stochastic diagrams has the
form of a Feynman diagram, apart from the fact that we have to
take into account that we are joining together two white random
noise fields many times.

As simple examples let us show how to derive the two-point
function in the zeroth order $\langle\,\varphi(\tau_{1},x_{1})
\varphi(\tau_{2},x_{2})\,\rangle^{(0)}_{\eta}$, and also the first
order correction to the scalar two-point-function given by
$\langle\,\varphi(\tau_{1},x_{1})
\varphi(\tau_{2},x_{2})\,\rangle^{(1)}_{\eta}$. Using the
Eq.(\ref{26}) and the Einstein relations we have
\begin{equation}
\langle\,\varphi(\tau_{1},x_{1})
\varphi(\tau_{2},x_{2})\,\rangle^{(0)}_{\eta} =2
\int_{0}^{min(\tau_{1},\tau_{2})}d\tau'\int_{\Omega}d^{d}x'\,
G(\tau_{1}-\tau',x_{1}-x')\,G(\tau_{2}-\tau',x_{2}-x'),
 \label{inc1}
\end{equation}
which is just Eq.(\ref{31}). For the first order correction we
get:
\begin{eqnarray}
&&\langle\,\varphi(X_{1})
\varphi(X_{2})\,\rangle^{(1)}_{\eta}=-\frac{\lambda}{3!}\langle
\int\,dX_{3}\int\,dX_{4}\Biggl(G(X_{1}-X_{4})G(X_{2}-X_{3})+\nonumber\\
&& G(X_{1}-X_{3})G(X_{2}-X_{4})\Bigg)\eta(X_{3})
\Biggl(\int\,dX_{5}\,G(X_{4}-X_{5})\eta(X_{5})\Bigg)^{3}\rangle_{\eta}.
\label{inc2}
\end{eqnarray}
where, for simplicity, we have introduced a compact notation:
\begin{equation}
\int_{0}^{\tau}d\tau\int_{\Omega}d^{d}x\equiv\int\,dX,
\end{equation}
and also $\varphi(\tau,x)\equiv\varphi(X)$ and finally
$\eta(\tau,x)\equiv\eta(X)$.

The process can be repeated and therefore the stochastic
quantization can be used as an alternative approach to describe
scalar quantum fields. We stress here that the stochastic
quantization is based in the fact that although one starts with
the system out of equilibrium, the Markovian Langevin equation
forces it into equilibrium. Moreover, when the thermodynamic
equilibrium is reached, the stochastic expectation values will
coincide with the Schwinger functions of the Euclidean field
theory.
\begin{figure}[ht]

\begin{picture}(100,100)
\put(30,30){{ \small$\varphi=$}}

\put(65,33){\line(1,0){20}}

\put(78,30){{ \small$\times$}}

\put(98,30){{ \small$+$}}

\put(125,33){\line(1,0){20}}

\put(145,33){\line(1,1){10}}

\put(145,33){\line(1,0){10}}

\put(145,33){\line(1,-1){10}}

\put(148,40){{ \small$\times$}}

\put(148,30){{ \small$\times$}}

\put(148,20){{ \small$\times$}}

\put(178,30){{ \small$+$}}

\put(205,33){\line(1,0){35}}

\put(225,33){\line(1,1){17}}


\put(225,33){\line(1,-1){15}}

\put(233,30){{ \small$\times$}}

\put(233,16){{ \small$\times$}}

\put(234,43){\line(1,0){10}}

\put(234,43){\line(0,1){10}}

\put(237,40){{ \small$\times$}}

\put(227,51){{ \small$\times$}}

\put(234,48){{ \small$\times$}}

\put(268,30){{ \small$+\;\;\;\;...$}}

\end{picture}

\caption[region] {Perturbative expansion for the scalar field
where crosses denote noise fields.}

\end{figure}

\begin{figure}[ht]

\begin{picture}(100,100)
\put(30,30){{ \small$\langle\,\varphi\,\varphi\,\rangle_{\eta}
=$}}

\put(95,33){\line(1,0){40}}

\put(108,30){{ \small$\times$}}

\put(90,20){{\small$\tau_1$}}

\put(135,20){{\small $\tau_2$}}

\put(103,0){{\small $(a)$}}

\put(155,30){{ \small$+$}}

\put(195,30){\line(1,0){70}}

\put(190,20){{\small$\tau_1$}}

\put(265,20){{\small $\tau_2$}}

\put(230,42){\circle{25}}

\put(222,51.5){{ \small$\times$}}

\put(245,27){{ \small$\times$}}

\put(224,0){{\small $(b)$}}

\put(285,30){{ \small$+$}}

\put(325,30){\line(1,0){70}}

\put(320,20){{\small$\tau_1$}}

\put(395,20){{\small $\tau_2$}}

\put(360,42){\circle{25}}

\put(352,51.5){{ \small$\times$}}

\put(335,27){{ \small$\times$}}

\put(354,0){{\small $(c)$}}

\end{picture}

\caption[region] {The corrections up to one-loop to the two-point
correlation function.}

\end{figure}
\newpage
We can represent Eq.(\ref{41}) graphically as figure $(1)$ (the
random noise field is represented by a cross). Using this
diagrammatical expansion, it is possible to show that the
two-point correlation function up to one-loop level is given by
figure $(2)$, where we represent the retarded Green function by a
line and the free two-point function by a crossed line. The rules
to obtain the algebraic values of the stochastic diagrams are
similar to the usual Feynman rules. For instance the
two-point function at one-loop level is given by\\
\begin{equation}
(b)=-\frac{\lambda}{2}\,\delta^d(k_1+k_2)\int\,d^dk\int_{0}^{\tau_{1}}\,d\tau\,
G(k_1;\tau_1-\tau)D(k;\tau,\tau)D(k_2;\tau_2,\tau).
\end{equation}
\begin{equation}
(c)=
-\frac{\lambda}{2}\,\delta^d(k_1+k_2)\int\,d^dk\int_{0}^{\tau_{2}}\,d\tau\,
G(k_2;\tau_2-\tau)D(k;\tau,\tau)D(k_1;\tau_1,\tau).\\
\end{equation}
A simple computation shows that we recover the correct equilibrium
result at equal asymptotic Markov parameters
($\tau_1=\tau_2\rightarrow \infty$):
\begin{equation}
(b)|_{\tau_1=\tau_2\rightarrow
\infty}=-\frac{\lambda}{2}\,\delta^d(k_1+k_2)\frac{1}{(k_2^2+m_0^2)}\frac{1}{(k_1^2+k_2^2+2m_0^2)}
\int\,d^dk\frac{1}{(k^2+m_0^2)}. \label{700}
\end{equation}

Obtaining the Schwinger functions in the asymptotic limit does not
guarantee that we gain a finite physical theory. The next step is
to implement a suitable regularization scheme. A crucial point to
find a satisfactory regularization scheme is to use one that
preserves the symmetries of the original model. In the stochastic
regularization method the symmetries of the physical theory is
maintained. There are in general two different ways to implement
the stochastic regularization. The first one is to start from a
Langevin equation with a memory kernel. It is known from the
literature \cite{zbern} that this method can at best only remove
two degrees of divergence. Another possibility is to smear only
the noise field in the probability functional \cite{bgz}
\cite{alfaro}:
\begin{equation}
\langle\,F[\varphi]\,\rangle_{\eta}=
\frac{\int\,[d\eta]F[\varphi]\exp\biggl[-\frac{1}{4} \int d^{d}x
\int d\tau\, \int
d\tau'\,\eta(\tau,x)K_\Lambda^{-1}\eta(\tau',x)\bigg]}
{\int\,[d\eta]\exp\biggl[-\frac{1}{4} \int d^{d}x \int d\tau\int
d\tau'\,\eta(\tau,x)K_\Lambda^{-1}\eta(\tau',x)\bigg]}, \label{36}
\end{equation}
where $K_\Lambda$ is a memory kernel. In this case we change the
Einstein relations of the noise field to:
\begin{equation}
\langle\, \eta(\tau,x)\,\eta(\tau',x')\,
\rangle_{\eta}\,=2\,K_{\Lambda}(\tau,\tau')\,\delta^{d}(x-x').
\label{29a}
\end{equation}
The smearing function should be chosen such that, when
$\Lambda\rightarrow \infty$:
\begin{equation}
\lim\limits_{\Lambda\rightarrow
\infty}K_\Lambda(\tau-\tau')=\delta(\tau-\tau'),
\end{equation}
recovering the usual theory.

Since the Langevin equation is unaffected by the stochastic
regularization, the physical field is the same as in the regularized
case. However, the zeroth-order two-point correlation function is
given by:
\begin{eqnarray}
&&D(k;\tau,\tau')=\nonumber\\
&&2\delta^{d}(k+k')\int_{0}^{\tau} ds\int_{0}^{\tau'}
ds'\,G(k;\tau-s)\,G(k;\tau'-s')K_\Lambda(s-s') =\nonumber\\
&&2\delta^{d}(k+k')\int_{0}^{\tau} ds\int_{0}^{\tau'}
ds'\exp\Bigl(-(\tau+\tau'-s-s')(k^2+m_0^2)\Bigr)K_\Lambda(s-s').
\end{eqnarray}
It is possible to prove that a necessary condition that the
regularization function $K_\Lambda$ should satisfy in order to
render the divergent loops finite is
$K_{\Lambda}(\tau)\mid_{\tau=0}=0$. The following series of
kernels obeying this condition were proposed:
\begin{equation}
K^{(n)}_{\Lambda}(\tau)=\frac{1}{2n!}\Lambda^2(\Lambda^2\mid\tau\mid)^n
\exp\bigl(-\Lambda^2\mid\tau\mid\bigr).      \label{k1}
\end{equation}
For the case $n=0$ we obtain, for the free two-point correlation
function:
\begin{equation}
\lim\limits_{\atop{\tau\rightarrow \infty}}D(k;\tau,\tau)=
\frac{\delta^{d}(k+k')}{(k^2+m_0^2)}\,\frac{\Lambda^2}{(\Lambda^2+k^2+m_0^2)}.
\end{equation}
Since the stochastic diagrams contains crossed lines in its loops,
we have that the ultraviolet divergences can be regularized
choosing an appropriate $n$. Note that it is possible to use a
different regulator of the type
$K_{\sigma}(\tau)=\frac{1}{2}\,\sigma\,\tau^{\sigma-1}$. This
regulator scheme is quite similar to the analytic regularization
of Bollini et al \cite{bol} and Speer \cite{speer}. The relation
between the cutoff regularization and the analytic regularization
procedure has been clarified by Kay \cite{kay} and also Svaiter
and Svaiter \cite{cas1} \cite{cas2} \cite{cas3} in a series of
papers studying the Casimir effect. In the next section we will
study a generalized Langevin equation with memory kernel and
colored noise.

\section{Generalized Langevin
equation with colored noise.}\

It is well known that the Langevin equation given by Eq.(\ref{23})
is only one particular choice in a large class of relaxation
equations. To make sure that the generalized Langevin equation can
also be used as a quantization tool, one must check that the
process converges in the asymptotic limit and also that converges
to the correct equilibrium distribution. Therefore, the aim of
this section is to discuss the $(\lambda\varphi^{4})_{d}$ scalar
field theory that appears if we start from a Langevin equation
with a memory kernel and also a colored noise. To proceed, let us
introduce a Langevin equation with a memory kernel given by
\begin{equation}
\frac{\partial}{\partial\tau}\varphi(\tau,x)
=-\int_{0}^{\tau}ds\,K_\Lambda(\tau-s)
\frac{\delta\,S}{\delta\,\varphi(x)}|_{\varphi(x)=\varphi(s,\,x)}+
\eta(\tau,x), \label{93}
\end{equation}
where the stochastic random field $\eta(\tau,x)$ satisfies the
modified Einstein relations.
\begin{equation}
\langle\,\eta(\tau,x)\,\eta(\tau',x')\,\rangle
_{\eta}=2K_{\Lambda}(|\tau-\tau'|)\,\delta^{d}(x-x'). \label{94}
\end{equation}
In this case where $K_{\Lambda}(|\tau-\tau'|)$ has a width in the
fictitious time, the description is Gaussian in spite of being
non-Markovian. For the free scalar field we have that the
generalized Langevin equation reads
\begin{equation}
\frac{\partial}{\partial\tau}\varphi(\tau,x)
=-\int_{0}^{\tau}ds\,K_\Lambda(\tau-s)
(-\Delta+m_{0}^{2})\varphi(s,x)+ \eta(\tau,x). \label{95}
\end{equation}
Using a Fourier representation for the scalar field and the random
noise field we get
\begin{equation}
\frac{\partial}{\partial\tau}\varphi(\tau,k)=
-(k^{2}+m_{0}^{2})\int_{0}^{\tau}ds\,K_{\Lambda}(\tau-s)
\varphi(s,k)+ \eta(\tau,k). \label{96}
\end{equation}
Following Ref. \cite{fox1}, we define the Laplace transform of the
memory kernel:
\begin{equation}
{K}(z)=\int_{0}^{\infty}d\tau\,K_{\Lambda}(\tau)\,e^{-z\tau}.
\label{n7}
\end{equation}
With the initial condition $\varphi(\tau,k)|_{\tau=0}=0$, the
solution of the Eq.(\ref{96}) reads:
\begin{equation}
\varphi(\tau,k)= \int_{0}^{\infty}d\tau'\,G(k,\tau-\tau')\,
\eta(\tau',k), \label{k2}
\end{equation}
where using the step function $\theta(\tau)$, the Green function
$G(k,\tau)$ is defined by:
\begin{equation}
G(k,\tau)\equiv\Omega(k,\tau)\,\theta(\tau).
\label{k3a}
\end{equation}
The $\Omega(k,\tau)$ function that appears in Eq.(\ref{k3a}) is
defined through its Laplace transform:
\begin{equation}
\Omega(k,\tau)=\biggl(z+(k^2+m_0^2){K}(z)\biggr)^{-1}.  \label{k5}
\end{equation}
It is easy to see that, in the limit
$K_{\Lambda}(\tau)\rightarrow\delta(\tau)$, we obtain the usual
result for the Green function. From Eq.(\ref{k2}) and the modified
Einstein relations, we get that the free scalar correlation
function $D(k;\tau,\tau')$ is given by:
\begin{eqnarray}
&&D(k;\tau,\tau')=\nonumber\\
&& = 2\delta^{d}(k+k')\int_{0}^{\infty} ds\int_{0}^{\infty}
ds'\,G(k,\tau-s)\,G(k,\tau'-s')\,K_\Lambda(\mid s-s' \mid) \nonumber\\
&& = 2\delta^{d}(k+k')\int_{0}^{\tau} ds\int_{0}^{\tau'}
ds'\,\Omega(k,\tau-s)\,\Omega(k,\tau'-s')\,K_\Lambda(\mid s-s'
\mid). \label{k4}
\end{eqnarray}
To proceed we have to write $D(k;\tau,\tau')$ in a simplified way.
Note that the double Laplace transform of the right hand side is
given by:
\begin{eqnarray}
&&\int_{0}^{\infty}d\tau\,e^{-z\tau}\int_{0}^{\infty}d\tau'\,e^{-z\tau'}\int_{0}^{\tau}
ds\int_{0}^{\tau'}
ds'\,\Omega(k,\tau-s)\,\Omega(k,\tau'-s')\,K_\Lambda(\mid
s-s' \mid)=\nonumber\\
&& = \Omega(k,z)\,\Omega(k,z') \int_{0}^{\infty}
ds\int_{0}^{\infty} ds'\,e^{-z's'}e^{-zs}\,K_\Lambda(\mid s-s'
\mid). \label{k6}
\end{eqnarray}
Now, with simple manipulations, we get:
\begin{equation}
\int_{0}^{\infty} ds\int_{0}^{\infty}
ds'\,e^{-z's'}e^{-zs}K_\Lambda(\mid s-s'
\mid)=\frac{{K}(z)+{K}(z')}{z+z'}. \label{k7}
\end{equation}
Therefore, we get the identity:
\begin{eqnarray}
&&\int_{0}^{\infty}d\tau\,e^{-z\tau}\int_{0}^{\infty}d\tau'\,e^{-z\tau'}\int_{0}^{\tau}
ds\int_{0}^{\tau'}
ds'\Omega(k,\tau-s)\,\Omega(k,\tau'-s')\,K_\Lambda(\mid s-s' \mid)=\nonumber\\
&& = \Omega(k,z)\,\Omega(k,z')\Biggl(\frac{K(z)+K(z')}{z+z'}\Biggr).
\label{k8}
\end{eqnarray}
Remembering Eq.(\ref{k5}), we can show that:
\begin{equation}
\Omega(k,z)\,\Omega(k,z')\Biggl(\frac{K(z)+K(z')}{z+z'}\Biggr)=\frac{1}{(k^2+m_0^2)}
\Biggl(\frac{\Omega(k,z)+\Omega(k,z')}{z+z'}-\Omega(k,z)\,\Omega(k,z')\Biggr).
\label{k9}
\end{equation}
So, in parallel with result Eq.(\ref{k7}), we finally obtain a
very simple expression for $D(k;\tau,\tau')$ in terms of
$\Omega(k,\tau)$. We have
\begin{equation}
D(k;\tau,\tau')= 2\,\frac{\delta^{d}(k+k')}{(k^2+m_0^2)}\,
\biggl(\Omega(k,\mid
\tau-\tau'\mid)-\Omega(k,\tau)\,\Omega(k,\tau')\biggr).
\label{k10}
\end{equation}
Now, we need an expression for our memory kernel in order to
investigate the convergence of Eq.(\ref{k10}). From Eq.(\ref{k1}),
we will have, for $n=0$:
\begin{equation}
K_\Lambda(\tau)=\frac{1}{2}\,\Lambda^2\,\exp\bigl(-\Lambda^2\mid\tau\mid\bigr).
\label{k11}
\end{equation}
Then, from Eq.(\ref{n7}), Eq.(\ref{k5}) and Eq.(\ref{k11}), and
applying the inverse Laplace transform, we will obtain the
following expression for the $\Omega$-function:
\begin{equation}
\Omega(k,\tau)=\Biggl(\frac{\Lambda^2}{\beta}\sin\biggl
(\frac{\beta\tau}{2}\biggr)+\cos\biggl(\frac{\beta\tau}{2}\biggr)\Biggr)\,
\exp\biggl(-\tau\frac{\,\Lambda^2}{2}\biggr).   \label{k12}
\end{equation}
where we have defined a real quantity $\beta$ given by:
\begin{equation}
\beta\equiv\Lambda\sqrt{2(k^2+m_0^2)-\Lambda^2}.     \label{k13}
\end{equation}

Similarly, we will have, for the Green function:
\begin{equation}
G(k,\tau)=\Biggl(\frac{\Lambda^2}{\beta}\sin\biggl(\frac{\beta\tau}{2}\biggr)
+\cos\biggl(\frac{\beta\tau}{2}\biggr)\Biggr)\,
\exp\biggl(-\tau\frac{\,\Lambda^2}{2}\biggr)\theta(\tau).
\label{k14}
\end{equation}
From the results above, it is easy to see that the free two-point function will be given by:
\begin{eqnarray}
&&D(k;\tau,\tau')=\nonumber\\
&& = 2\,\frac{\delta^{d}(k+k')}{(k^2+m_0^2)}\,
\Biggl[\Biggl(\frac{\Lambda^2}{\beta}\,\sin\biggl(\frac{\beta(\tau-\tau')}{2}\biggr)
+\,\cos\biggl(\frac{\beta(\tau-\tau')}{2}\biggr)\Biggr)
\,\exp\biggl(-\frac{\,\Lambda^2}{2}\mid(\tau-\tau')\mid\biggr)  \nonumber\\
&&-\Biggl(\frac{\Lambda^2}{\beta}\,\sin\biggl(\frac{\beta\tau}{2}\biggr)+\,
\cos\biggl(\frac{\beta\tau}{2}\biggr)\Biggr)\,
\Biggl(\frac{\Lambda^2}{\beta}\,\sin\biggl(\frac{\beta\tau'}{2}\biggr)+\,
\cos\biggl(\frac{\beta\tau'}{2}\biggr)\Biggr)\,
\exp\biggl(-\frac{\,\Lambda^2}{2}(\tau+\tau')\biggr)\Biggr].
\label{k15}
\end{eqnarray}
For $\tau=\tau'$, we get:
\begin{equation}
D(k;\tau,\tau)= 2\,\frac{\delta^{d}(k+k')}{(k^2+m_0^2)}\,
\Biggl(1-\Biggl(\frac{\Lambda^2}{\beta}\,\sin\biggl(\frac{\beta\tau}{2}\biggr)+\,
\cos\biggl(\frac{\beta\tau}{2}\biggr)\Biggr)^2\exp\bigl(-\Lambda^2\,\tau\bigr)\Biggr)\,
. \label{k16}
\end{equation}
So, in the limit $\tau\rightarrow\infty$, we obtain the following result:
\begin{equation}
D(k;\tau,\tau)= 2\,\frac{\delta^{d}(k+k')}{(k^2+m_0^2)}.
\label{k17}
\end{equation}
which, as we see, does not present an improved ultraviolet
behavior. In fact, this equation is much similar to the usual
equilibrium result, up to a constant. This is a first evidence
that the ultraviolet divergences appearing in the perturbative
series when we consider the self-interacting theory may not be
regularized. Now, let us study the self-interaction
$(\lambda\varphi^{4})_{d}$ scalar field theory within this
non-Markovian approach. Now, the Langevin equation reads:
\begin{equation}
\frac{\partial}{\partial\tau}\varphi(\tau,x)
=-\int_{0}^{\tau}ds\,K_\Lambda(\tau-s)
\Bigl((-\Delta+m_{0}^{2})\varphi(s,x)+\frac{\lambda}{3!}\varphi^{3}(s,x)\Bigr)+
\eta(\tau,x). \label{s1}
\end{equation}
We can solve this equation by iteration as before. Then, after
equating terms with equal powers in $\lambda$, we get:
\begin{equation}
\Biggl[\frac{\partial}{\partial\tau}+(-\Delta_{x}+m^{2}_{0}\,)
\Bigg]\varphi^{(0)}(\tau,x)=\eta(\tau,x), \label{s2}
\end{equation}
\begin{equation}
\Biggl[\frac{\partial}{\partial\tau}+(-\Delta_{x}+m^{2}_{0}\,)
\Bigg]\varphi^{(1)}(\tau,x)=-\frac{1}{3!}\int_{0}^{\tau}ds\,K_\Lambda(\tau-s)
\left(\varphi^{(0)}(s,x)\right)^{3}, \label{s3}
\end{equation}
\begin{equation}
\Biggl[\frac{\partial}{\partial\tau}+(-\Delta_{x}+m^{2}_{0}\,)
\Bigg]\varphi^{(2)}(\tau,x)=-\frac{1}{2!}\int_{0}^{\tau}ds\,K_\Lambda(\tau-s)
\left(\varphi^{(0)}(s,x)\right)^{2}\varphi^{(1)}(s,x), \label{s4}
\end{equation}
and so on. The solutions of the first two equations can be written
as:
\begin{equation}
\varphi^{(0)}(\tau,x)=\int_{0}^{\tau}d\tau'\int_{\Omega}d^{d}x'\,
G(\tau-\tau',x-x')\eta(\tau',x'). \label{s5}
\end{equation}
and
\begin{eqnarray}
\varphi^{(1)}(\tau,x)&=&
-\frac{1}{3!}\int_{0}^{\tau}d\tau_{1}\int_{\Omega}d^{d}x_{1}\,
G(\tau-\tau_{1},x-x_{1})\int_{0}^{\tau_{1}}ds\,K_\Lambda(\tau_{1}-s)\nonumber \\
&&\left(\int_{0}^{\tau_{1}}d\tau'\int_{\Omega}d^{d}x'\,
G(\tau_{1}-\tau',x_{1}-x')\eta(\tau',x') \right)^{3}, \label{s6}
\end{eqnarray}
with the Green function given by Eq.(\ref{k14}).

For the $n$-point correlation functions, the perturbation theory
will be similar to the Markovian case, except that for each vertex
in the stochastic diagram there is a memory kernel associated. For
instance, the one-loop correction for the two-point function is
given by ($(b)$ and $(c)$ given by figure $(2)$):
\begin{equation}
(b)=-\frac{\lambda}{2}\,\delta^d(k_1+k_2)\int\,d^dk\int_{0}^{\tau_{1}}\,d\tau\int_{0}^{\tau}\,ds\,
G(k_1,\tau_1-\tau)D(k;s,s)D(k_2;\tau_2,s)K_\Lambda(\mid\tau-s\mid).
\label{s7}
\end{equation}
\begin{equation}
(c)=
-\frac{\lambda}{2}\,\delta^d(k_1+k_2)\int\,d^dk\int_{0}^{\tau_{2}}\,d\tau\int_{0}^{\tau}\,ds\,
G(k_2,\tau_2-\tau)D(k;s,s)D(k_1;\tau_1,s)K_\Lambda(\mid\tau-s\mid).\label{s8}\\
\end{equation}
From Eq.(\ref{k14}) and Eq.(\ref{k15}) we may split Eq.(\ref{s7})
in four parts as:
\begin{equation}
(b)=
-\frac{\lambda}{2}\,\delta^d(k_1+k_2)\int\,d^dk\int_{0}^{\tau_{1}}\,d\tau\int_{0}^{\tau}\,ds\,
\Bigl(I_1+I_2+I_3+I_4\Bigr),      \label{s9}
\end{equation}
where:
\begin{equation}
I_1\equiv\int_{0}^{\tau_{1}}\,d\tau\int_{0}^{\tau}\,ds\,\Omega(k_1,\tau_1-\tau)\,
\Omega(k_2,\tau_2-s)\,\Omega(k,0)\,K_\Lambda(\mid\tau-s\mid),
\end{equation}
\begin{equation}
I_2\equiv\int_{0}^{\tau_{1}}\,d\tau\int_{0}^{\tau}\,ds\,\Omega(k_1,\tau_1-\tau)\,
\Omega(k_2,\tau_2)\,\Omega(k_2,s)\,\Omega(k,0)\,K_\Lambda(\mid\tau-s\mid),
\end{equation}
\begin{equation}
I_3\equiv\int_{0}^{\tau_{1}}\,d\tau\int_{0}^{\tau}\,ds\,\Omega(k_1,\tau_1-\tau)\,
\Omega(k_{2},\tau_{2}-s)\,\Omega^2(k,s)\,K_\Lambda(\mid\tau-s\mid),
\end{equation}
\begin{equation}
I_4\equiv\int_{0}^{\tau_{1}}\,d\tau\int_{0}^{\tau}\,ds\,\Omega(k_1,\tau_1-\tau)\,
\Omega(k_2,\tau_2)\,\Omega(k_2,s)\,\Omega^2(k,s)\,K_\Lambda(\mid\tau-s\mid).
\end{equation}
Using simple calculations, it is easy to show that the terms
$I_2$, $I_3$ and $I_4$ have a decaying oscillatory regime for
$\tau_1=\tau_2\rightarrow\infty$. So, we only have to consider the
first term. It reads:
\begin{eqnarray}
&&I_1=\nonumber\\
&&
=\frac{\Lambda^2}{2}\,\exp\biggl(-\frac{\Lambda^2}{2}(\tau_{1}+\tau_{2})\biggr)
\int_{0}^{\tau_{1}}\,d\tau\Biggl(\frac{\Lambda^2}{\beta_{1}}
\sin\biggl(\frac{\beta_{1}(\tau_{1}-\tau)}{2}\biggr)
+\cos\biggl(\frac{\beta_{1}(\tau_{1}-\tau)}{2}\biggr)\Biggr)\nonumber\\
&&\exp\biggl(-\frac{\Lambda^2}{2}\tau\biggr)
\int_{0}^{\tau}ds\,\Biggl(\frac{\Lambda^2}{\beta_{2}}
\sin\biggl(\frac{\beta_{2}(\tau_{2}-s)}{2}\biggr)+
\cos\biggl(\frac{\beta_{2}(\tau_{2}-s)}{2}\biggr)\Biggr)\,
\exp\biggl(\frac{3\Lambda^2}{2}s\biggr).
\label{s10}
\end{eqnarray}
After simple manipulations \cite{grads} and a tedious algebra, we
obtain the final result, in the limit
$\tau_1=\tau_2\rightarrow\infty$:
\begin{equation}
(b)|_{\tau_1=\tau_2\rightarrow\infty}
=-2\lambda\,\frac{\delta^d(k_1+k_2)}{(k_2^2+m_0^2)}\,f(\Lambda;\beta_{1},\beta_{2})
\int\,d^dk\,\frac{1}{(k^2+m_0^2)},    \label{s11}
\end{equation}
where:
\begin{equation}
f(\Lambda;\beta_{1},\beta_{2})\equiv
\Biggl(\frac{\bigl(72\,\Lambda^{4}+9\,\beta_{1}^{2}-\beta_{2}^{2}\bigr)\,
\Lambda^4+\beta_{2}^{2}\bigl(\beta_{1}^{2}-\beta_{2}^{2}\bigr)}
{\bigl(4\Lambda^{4}+(\beta_{1}-\beta_{2})^2\bigr)\bigl(4\Lambda^{4}+(\beta_{1}-\beta_{2})^2\bigr)}\Biggr)
\Biggl(\frac{2\,\Lambda^2}{9\Lambda^{4}+\beta_{2}^{2}}\Biggr),
\label{s12}
\end{equation}
and $\beta_i=\Lambda\sqrt{2(k_{i}^2+m_0^2)-\Lambda^2}; i=1,2$.
So, we see that, although we obtain convergence in the asymptotic
limit of the fictitious parameter $\tau$, we do not get a
regularized theory. In fact, up to polynomials of $\Lambda$, the
result is much similar to the usual equilibrium result,
Eq.(\ref{700}). Since the usual stochastic regularization requires
the smearing of only the $\eta$ probability functional, leaving
the Langevin equation alone, a natural question that arises is why
a theory with a colored, internal noise does not lead to finite
results in perturbation theory. Let us try to answer this question
within a Fokker-Planck analysis.

\section{The Fokker-Planck approach}

Although in this paper we study the stochastic perturbation theory
using the Langevin equation approach, to understand our results it
is more suitable to work within the Fokker-Planck formulation. As
we know, correlation functions are introduced as averages over
$\eta$:
\begin{eqnarray}
&&\langle\,\varphi
(\tau_{1},x_{1})\varphi(\tau_{2},x_{2})\cdots\varphi(\tau_{n},x_{n})\,\rangle_{\eta}=\nonumber\\
&&{\cal{N}} \int\,[d\eta]\varphi(\tau_{1},x_{1})
\varphi(\tau_{2},x_{2})\cdots\varphi(\tau_{n},x_{n})
\exp\biggl(-\frac{1}{4} \int d^{d}x \int
d\tau\,\eta^{2}(\tau,x)\bigg), \label{s12}
\end{eqnarray}
where ${\cal{N}}$ is given by
\begin{equation}
{\cal{N}}^{-1}=\int\,[d\eta] \exp\biggl(-\frac{1}{4} \int d^{d}x
\int d\tau\,\eta^{2}(\tau,x)\bigg).
\end{equation}

An alternative way to write this average is to introduce the
probability density $P[\varphi,\tau]$, which is defined as
\cite{ili}:
\begin{equation}
P[\varphi,\tau]\equiv\int\,[d\eta]\exp\biggl(-\frac{1}{4} \int
d^{d}x \int
d\tau'\,\eta^{2}(\tau',x)\bigg)\,\prod\limits_{y}\delta(\varphi(y)-\varphi(\tau,y)).\label{s13}
\end{equation}
where, for simplicity, we have absorbed the factor ${\cal{N}}$ in
the original functional measure. In terms of $P[\varphi,\tau]$,
the equal time correlation functions will read:
\begin{equation}
\langle\,\varphi
(\tau_{1},x_{1})\varphi(\tau_{2},x_{2})\cdots\varphi(\tau_{n},x_{n})\,\rangle_{\eta}=
\int\,[d\varphi]\,\varphi(\tau_{1},x_{1})
\varphi(\tau_{2},x_{2})\cdots\varphi(\tau_{n},x_{n})P[\varphi,\tau].
\label{s14}
\end{equation}
The probability density $P[\varphi,\tau]$ satisfies the
Fokker-Planck equation \cite{fokker}:
\begin{equation}
\frac{\partial}{\partial\tau}P[\varphi,\tau]=\int\,d^{d}x
\frac{\delta}{\delta\,\varphi(x)}\Biggl(\frac{\delta}{\delta\,\varphi(x)}+
\frac{\delta\,S_0}{\delta\,\varphi(x)}\Biggr)P[\varphi,\tau],
\label{s15}
\end{equation}
with the initial condition:
\begin{equation}
P[\varphi,0] = \prod\limits_{y}\delta(\varphi(y)).
\end{equation}
The stochastic quantization says that we shall have:
\begin{equation}
w.\lim\limits_{\tau\rightarrow\infty}P[\varphi,\tau] =
\frac{\exp\bigl(-S[\varphi]\bigr)}
{\int\,[d\varphi]\exp\bigl(-S[\varphi]\bigr)},
 \label{s16}
\end{equation}
where the limit is supposed to be taken "weakly" in the sense of
the reference \cite{ili}. Again following the former reference,
for the perturbation theory, we may split the action in two parts:
\begin{equation}
S=S_{0}+\lambda\,S_{I}. \label{s17}
\end{equation}

The Green functions of the theory are computed as power series in
$\lambda$. Similarly, we may expand the probability density as:
\begin{equation}
P[\varphi,\tau]=\sum\limits_{0}^{\infty}P_{k}[\varphi,\tau].\label{s18}
\end{equation}
So, the $n$-point correlation function in the $k$th order of
perturbation theory will be:
\begin{equation}
\langle\,\varphi
(\tau_{1},x_{1})\varphi(\tau_{2},x_{2})\cdots\varphi(\tau_{n},x_{n})\,\rangle_{\eta}=
\lambda^{k}\int\,[d\varphi]\,\varphi(\tau_{1},x_{1})
\varphi(\tau_{2},x_{2})\cdots\varphi(\tau_{n},x_{n})P_{k}[\varphi,\tau].
\label{s19}
\end{equation}
The Fokker-Planck equation becomes:
\begin{eqnarray}
\frac{\partial}{\partial\tau}P_{k}(\varphi,\tau)&=&\int\,d^{d}x
\frac{\delta}{\delta\,\varphi(x)}\Biggl(\frac{\delta}{\delta\,\varphi(x)}+
\frac{\delta\,S_0}{\delta\,\varphi(x)}\Biggr)P_{k}(\varphi,\tau)+\nonumber\\
&& + \int\,d^{d}x\frac{\delta}{\delta\,\varphi(x)}
\frac{\delta\,S_{I}}{\delta\,\varphi(x)}P_{k-1}(\varphi,\tau),
 \label{s20}
\end{eqnarray}
with the initial conditions:
\begin{equation}
P_0[\varphi,0] = \prod\limits_{y}\delta(\varphi(y)),
\end{equation}
\begin{equation}
P_{k}[\varphi,0] = 0,\,\,k=1,2,\cdots
\end{equation}
We may write a formal solution to the Fokker-Planck equation as:
\begin{equation}
P_{k}[\varphi,\tau]=\int\,[d\varphi']\,\int_{0}^{\tau}\,d\tau'\,{\cal{D}}_0[\varphi,\varphi',\tau-\tau']\,
\int\,dx\,\frac{\delta}{\delta\,\varphi'(x)}
\frac{\delta\,S_{I}}{\delta\,\varphi'(x)}P_{k-1}(\varphi',\tau'),\label{s21}
\end{equation}
where the Green functional of the free Fokker-Planck equation
satisfies:
\begin{eqnarray}
\frac{\partial}{\partial\tau}{\cal{D}}_0[\varphi,\varphi',\tau-\tau']&=&
\prod\limits_{x}\delta(\varphi(x)-\varphi'(x))\,\delta(\tau-\tau')+\nonumber\\
&& +
\int\,d^{d}x\frac{\delta}{\delta\,\varphi(x)}\Biggl(\frac{\delta}{\delta\,\varphi(x)}+
\frac{\delta\,S_0}{\delta\,\varphi(x)}\Biggr){\cal{D}}_0[\varphi,\varphi',\tau-\tau'],
\label{s22}
\end{eqnarray}
with the boundary condition:
\begin{equation}
{\cal{D}}_0[\varphi,\varphi',\tau-\tau']|_{\tau=\tau'_{+}}=\prod\limits_{x}\delta(\varphi(x)-\varphi'(x)).
\end{equation}
For our case, the above equations should be modified in order to
take into account the presence of the memory kernel. For instance,
the free Fokker-Planck equation will be:
\begin{equation}
\frac{\partial}{\partial\tau}P_0[\varphi,\tau]=\int\,d^{d}x
\int\,d^{d}y\,
\frac{\delta}{\delta\,\varphi(x)}\Biggl(\frac{\delta}{\delta\,\varphi(y)}M(x-y,\tau)+
\frac{\delta}{\delta\,\varphi(y)}N[\varphi\,;\tau,x]\Biggr)P_0[\varphi,\tau],
\label{s23}
\end{equation}
where:
\begin{equation}
M(x-y,\tau)\equiv\int_{0}^{\tau}\,ds\,K_{\Lambda}(\tau-s)\,G(\tau-s,x-y),
\end{equation}
and
\begin{equation}
N[\varphi\,;\tau,x]\equiv\delta(x-y)\int_{0}^{\tau}\,ds\,K_{\Lambda}(\tau-s)S_{0}\mid_{\varphi(x)=\varphi(s,x)}.
\end{equation}
Now, we can easily see why we required earlier that $\beta$,
defined in Eq.(\ref{k13}), should be real. Remembering the
familiar transformation,
$P[\varphi,\tau]=\psi[\varphi,\tau]\exp\Bigl(-\frac{\hat{S}[\varphi]}{2}\Bigr)$,
with $\hat{S}\equiv M^{-1}N$ ($M=M(x-y,\tau)$,
$N=N[\varphi\,;\tau,x]$), we can recast Eq.(\ref{s23}) into the
Schroedinger type equation:
\begin{equation}
\dot{\psi}=-2{\cal{H}}\psi,
\end{equation}
where ${\cal{H}}$ is the Fokker-Planck Hamiltonian:
\begin{equation}
{\cal{H}}\equiv\frac{1}{2} \int\,d^{d}x\int\,d^{d}y\,
Q(x)M(x-y,\tau)Q(y), \label{ss}
\end{equation}
where:
\begin{equation}
Q(x)\equiv\hat{\Pi}(x)+ \frac{1}{2}\Bigl[\hat{\Pi}(x),\hat{S}\Bigr],
\end{equation}
and $\hat{\Pi}(x)\equiv -i \frac{\delta}{\delta\varphi(x)}$. We
stress here that this derivation is purely formal.
Now, let us study the function $M(x-y,\tau)$. In the Fourier
space, it is given by:
\begin{eqnarray}
M(k,\tau)&=&\frac{\,\Lambda^{2}}{2}\,\frac{1}{9\,\Lambda^{4}+\beta^{2}}\biggl\{8\,\Lambda^{2}-\nonumber\\
&& 4\,\exp\Bigl(-\frac{3\Lambda^{2}}{2}\tau\Bigr)
\biggl[2\Lambda^{2}\cos\Bigl(\frac{\beta\tau}{2}\Bigr)+
\biggl(\frac{3\Lambda^{4}}{2\beta}-\frac{\beta}{2}\biggr)\sin\Bigl(\frac{\beta\tau}{2}\Bigr)\biggr]\biggr\},
\label{g}
\end{eqnarray}
where we have used Eq.(\ref{k11}) and Eq.(\ref{k14}). So, we see
that, in the limit $\tau\rightarrow\infty$, the $M$-function will be
positive definite which, according to Eq.(\ref{ss}), guarantees that
the Fokker-Planck Hamiltonian be positive definite as well. In other
words, the real part of its eigenvalues will be positive. But what
happens if $\beta$ is a purely imaginary number? As for the Green
function, we will have:
\begin{equation}
G(k,\tau)=\Biggl(\frac{\Lambda^2}{\beta}\sinh\biggl(\frac{\beta\tau}{2}\biggr)
+\cosh\biggl(\frac{\beta\tau}{2}\biggr)\Biggr)\,
\exp\biggl(-\tau\frac{\,\Lambda^2}{2}\biggr)\theta(\tau).
\label{dg}
\end{equation}
So, from Eq.(\ref{g}) and Eq.(\ref{dg}), it is easy to see that
$M(k,\tau)$ may not be positive definite and, therefore, neither
will be ${\cal{H}}$. Then, there is no guarantee that the real
part of the eigenvalues of the latter will be positive. In other
words, we cannot assure that, in the limit
$\tau\rightarrow\infty$, the system will reach its ground state,
i.e., that it converges to an equilibrium.

Now, let us study the solution to Eq.(\ref{s23}). It will be of the
form:
\begin{eqnarray}
P_0[\varphi,\tau]&=&\int\,[d\eta]\prod\limits_{y}\delta(\varphi(y)-\varphi(\tau,y))\nonumber\\
&& \exp\biggl(-\frac{1}{4}\int d^{d}x \int
d\tau''\int\,d\tau'\,\eta(\tau'',x)\,K_{\Lambda}^{-1}(\tau''-\tau')\,\eta(\tau',x)\bigg),
\label{s24}
\end{eqnarray}
where $\varphi(\tau,y)$ satisfies Eq.(\ref{k2}). We can rewrite
this last equation in the form:
\begin{eqnarray}
P_0[\varphi,\tau]&=&\int\,[d\eta]\int\,[d\xi]\exp\biggl(i\int\,d^{d}x\,\xi\Bigl
(\varphi(x)-\varphi(\tau,x)\Bigr)\biggr)\nonumber\\
&& \exp\biggl(-\frac{1}{4}\int d^{d}x \int
d\tau''\int\,d\tau'\,\eta(\tau'',x)\,K_{\Lambda}^{-1}(\tau''-\tau')\,\eta(\tau',x)\bigg).
\label{s25}
\end{eqnarray}
Using Eq.(\ref{k2}) and doing the $\eta$, $\xi$ functional
integrations, we obtain:
\begin{equation}
P_0[\varphi,\tau]=N_0^{-1}
\exp\biggl(-\frac{1}{2}\int\,d^{d}x\,\int\,d^{d}x'\varphi(x)\,D^{-1}\,(\tau,x;\tau,x')\varphi(x')\bigg),
\label{s26}
\end{equation}
where the $D$-function is just the free two-point correlation
function and $N_0^{-1}$ is a normalization factor. In momentum
space, we get, for $\tau\rightarrow\infty$:
\begin{equation}
P_0[\varphi,\tau]=N_0^{-1}
\exp\biggl(-\frac{1}{4}\int\,d^{d}k\,\varphi(k)(k^2+m_0^2)\varphi(-k)\bigg).
\label{s27}
\end{equation}
As we can see, this result resembles the usual equilibrium result,
up to a constant, in contrast with the result obtained by Breit et
al \cite{bgz}. Now, proceeding with similar steps (see reference
\cite{ili}), it is possible to show that the Green functional for
the free Fokker-Planck equation is given by ($\Delta_0^{-1}$ is a
normalization factor):
\begin{equation}
{\cal{D}}_0[\varphi,\varphi',\tau]=\Delta_0^{-1}
\exp\biggl(-\frac{1}{2}\int\,d^{d}k\,\varphi_i(k)D_{ij}(\tau)\varphi_j(-k)\bigg),
\label{s28}
\end{equation}
where:
\begin{equation}
\varphi_i\equiv\pmatrix{\varphi(k)&\varphi\,'(k)\cr},
\end{equation}
and the elements of the matrix $D_{ij}$ are given by:
\begin{equation}
D_{ij}(\tau)\equiv\pmatrix{\hfill
D(k;\tau,\tau)&-D(k;\tau,\tau)\,G(k,\tau)\hfill\cr \hfill
-D(k;\tau,\tau)\,G(k,\tau)&D(k;\tau,\tau)\,G(k,\tau)\,G(k,\tau)\hfill\cr}.
\end{equation}

It is easy to verify that, in the limit $\tau\rightarrow\infty$,
$P_0$ and ${\cal{D}}_0$ will satisfy, up to constants, similar
relations as obtained by Floratos and Iliopoulos \cite{ili}. This
behavior can be understood if we notice the similar structure
between the Green function of the usual Parisi-Wu scheme and our
non-Markovian approach. The same is true for both free two-point
correlation functions. However, for massless scalar theories,
those estimations in \cite{ili} decay as inverse power of $\tau$.
In our approach, even in this massless situation, an exponential
behavior is found. Therefore, in the limit
$\tau\rightarrow\infty$, we get an improved convergence.


 From all of these results we expect to obtain
probability functions in the perturbation theory that are very
close to those usual equilibrium probability densities. Therefore
the theory will not be regularized. As remarked on \cite{fox1},
nothing is lost by using a non-Markovian description in place of a
Markov description, as long as it is Gaussian.

\newpage

\section{Conclusions}

It is well known that the Langevin equation given by Eq.(\ref{23})
is only one particular choice in a large class of relaxation
equations. The aim of this paper was to investigate if the
Parisi-Wu quantization method can be extended assuming a Langevin
equation with a memory kernel with the modified Einstein
relations. Therefore, in this paper we discussed the stochastic
quantization for massive self-interacting scalar field, first
assuming a Markovian Langevin equation. Then, we studied the
$(\lambda\varphi^{4})$ theory at the one-loop level introducing a
non-Markovian Langevin equation and examine the interacting field
theory that appears in the asymptotic limit of the non-Markovian
process.

To make sure that the first modification can be used, one must
first check that the system evolves to the equilibrium in the
asymptotic limit. Second we have to show that converges to the
correct equilibrium distribution. We proved that although the
system evolves to equilibrium, we obtain a non-regularized theory.
In contrast with dimensional regularization \cite {dim1} \cite
{dim2} \cite {dim3} \cite {dim4} and also other regularization
procedures, a remarkable property of the stochastic regularization
is that it preserves all the symmetries of the original theory.
With our results, it is easy to see that the system described by
this non-Markovian Langevin equation converges, but the virtues of
the stochastic regularization that we hoped to appear in this
framework are lost.

A natural continuation of this paper is to discuss the stochastic
quantization of bosonic and femionic fields in general Riemannian
spaces. A different application of the stochastic quantization is
to discuss interacting field theory in the presence of macroscopic
structures. It is well known that performing the weak-coupling
perturbative expansion, to renormalize the interacting field
theory, we have to introduce not only the usual bulk counterterms,
but also surface counterterms. This can be done at the one-loop
level at zero and finite temperature \cite{boun1} \cite{boun2}
\cite{boun3}. To extend the calculation to high-order loops will
appear overlapping divergences. A natural question is how to
implement the stochastic quantization in systems with macroscopic
structures. This subject is under investigation by the authors.

\section{Acknowledgements}

 One of the authors (G. Menezes) would like to thank D. Soares-Pinto
and G. Flores Hidalgo for useful discussions. This paper was
supported by the Conselho Nacional de Desenvolvimento Cientifico e
Tecnol{\'o}gico of Brazil (CNPq).

\end{document}